\documentclass[aps,prl,twocolumn,superscriptaddress,nobibnotes,amsmath,amssymb,floatfix,longbibliography]{revtex4-2}

\usepackage[utf8]{inputenc}
\usepackage[T1]{fontenc}
\usepackage{hyperref}
\usepackage{url}

\usepackage{amsmath}
\usepackage{amsfonts}
\usepackage{amssymb}
\usepackage{mathtools}
\usepackage{lmodern}
\usepackage{upgreek}
\usepackage{ragged2e}
\usepackage[version=3]{mhchem}
\usepackage[free-standing-units]{siunitx}

\usepackage{tabularx}
\usepackage{booktabs}
\usepackage{placeins,afterpage}
\usepackage{subcaption}
\DeclareCaptionLabelFormat{bold}{\textbf{(#2)}}
\DeclareCaptionJustification{justified}{\justifying}
\usepackage{epstopdf}
\usepackage{graphicx}
\usepackage{varwidth}
\usepackage{xcolor}

\begin{document}
\title{High-gain far-detuned nonlinear frequency conversion in optical fibers:\\ intramodal vs. intermodal processes}

\def \ICB{Laboratoire Interdisciplinaire Carnot de Bourgogne (ICB), UMR6303 CNRS-Université Bourgogne Franche-Comté, 21000 Dijon, France}
\def \WUST{Department of Optics and Photonics, Wrocław University of Science and Technology, Wybrzeże Wyspiańskiego 27, 50-370 Wrocław, Poland}

\author{Karolina~Stefańska}
\affiliation{\ICB}
\affiliation{\WUST}

\author{Pierre~Béjot}
\affiliation{\ICB}

\author{Julien~Fatome}
\affiliation{\ICB}

\author{Guy~Millot}
\affiliation{\ICB}

\author{Karol Lech Tarnowski}
\affiliation{\WUST}

\author{Bertrand~Kibler}
\email[]{bertrand.kibler@u-bourgogne.fr}
\affiliation{\ICB}

\begin{abstract}
We present theoretical and experimental evidence of high-gain far-detuned nonlinear frequency conversion, extending towards both the visible and the mid-infrared, in a few-mode graded-index silica fiber pumped at \SI{1.064}{\micro\meter}, and more specifically achieving gains of hundreds of dB per meter below  \SI{0.65}{\micro\meter} and beyond \SI{3.5}{\micro\meter}. Our findings highlight the potential of graded-index fibers in terms of strong modal confinement over an ultrabroad spectral range for enabling high-gain wavelength conversion. Such advancements require an accurate interpretation of intramodal and intermodal four-wave mixing processes.
\end{abstract}

\maketitle
\section*{Introduction}
In recent years, increasing the number of guided modes within an optical fiber has introduced a new degree of freedom for investigating rich and complex spatiotemporal dynamics as well as novel phase-matched frequency conversion processes.
The interactions of multiple modes have led to the observation of remarkable phenomena, including geometric parametric instability, beam self-cleaning, multimode solitons, and discretized conical emission, to name a few~\cite{wright2015,krupa2016,krupa2017,renninger2013,stef2023}.
These phenomena are intricately dependent on the opto-geometric parameters of the multimode fibers (MMFs).
Distinct guiding properties can be obtained through step-index and graded-index profiles, thus resulting in either few-mode or largely multimode behavior.
As an example, graded-index (GRIN) multimode fibers exhibit the feature of nearly equally-spaced effective refractive indices of guided modes, thereby creating favored conditions for the emergence of new phase-matched processes through the self-imaging effect.
In particular, collective spectral sidebands (associated with numerous intermodal four-wave mixing combinations) can emerge on both sides of the pump wavelength~\cite{krupa2016}, a phenomenon known as the geometric parametric instability (GPI).
When considering a few-mode fiber, the number of intermodal couplings is significantly reduced and easier to analyze.
Therefore, any newly generated frequency can be investigated through the phase-matching conditions of each potentially involved four-wave mixing (FWM) process~\cite{nazemosadat2016,perrey2019}.

Intermodal FWM and GPI in MMFs have recently been the subject of detailed investigations, with a particular focus on  the spectral peaks emerging on the short-wavelength side of the pump laser (i.e., in the near-infrared or visible region) in fibers with relatively large core diameter (50-\SI{100}{\micro\meter})~\cite{krupa2016,krupa2016spatiotemporal,lopez2019,bendahmane2021}. To the best of our knowledge, the possibility of generating such sidebands far from the pump wavelength in the mid-infrared (mid-IR) region has not been investigated experimentally so far. This is mainly due to the limited transparency window of silica fibers (\SI{0.4} - \SI{2.4}{\micro\meter}) commonly accepted, which prevents their practical use in the mid-IR. Nevertheless, it is worth mentioning that supercontinuum generation has been previously demonstrated up to \SI{3.0} - \SI{3.6}{\micro\meter} in short meter- (centimeter-)long segments of single-mode germanium-doped silica fibers \cite{xia2007,yin2016} by means of ultrashort pulse propagation and Raman scattering in cascaded fiber systems. A few experimental studies in the continuous-wave pumping limit have also explored the possibility of achieving direct FWM-based frequency conversion to the mid-IR in endlessly single-mode silica photonic crystal fibers, which enable precise engineering of dispersion properties and phasematching \cite{Jauregui2012,Herzog2012}. Typically low conversion efficiencies (0.1-1.0\%) estimated around \SI{3.0}{\micro\meter} were obtained due to strong fiber losses, without any measurement of both the mid-IR spectrum and degenerate FWM gain.

In the present work, we investigate far-detuned nonlinear frequency conversion through intramodal and intermodal phase-matchings in a few-mode fiber both theoretically and experimentally.
In particular, we unveil the potential of graded-index fibers in terms of strong modal confinement over an ultrabroad spectral range, thereby achieving high-gain wavelength conversion in very short fiber segments. Our study requires a detailed analysis of the full frequency-dependence of propagation constants for the distinct modes as well as their effective mode overlaps. Although our results confirm that efficient, broadband wavelength conversion in MMFs is a relevant alternative to single-mode fibers, it can still benefit from further development through careful engineering of the index profile and the use of infrared glasses.

    \begin{figure*}[ht!]\centering\includegraphics[width=0.8\linewidth]{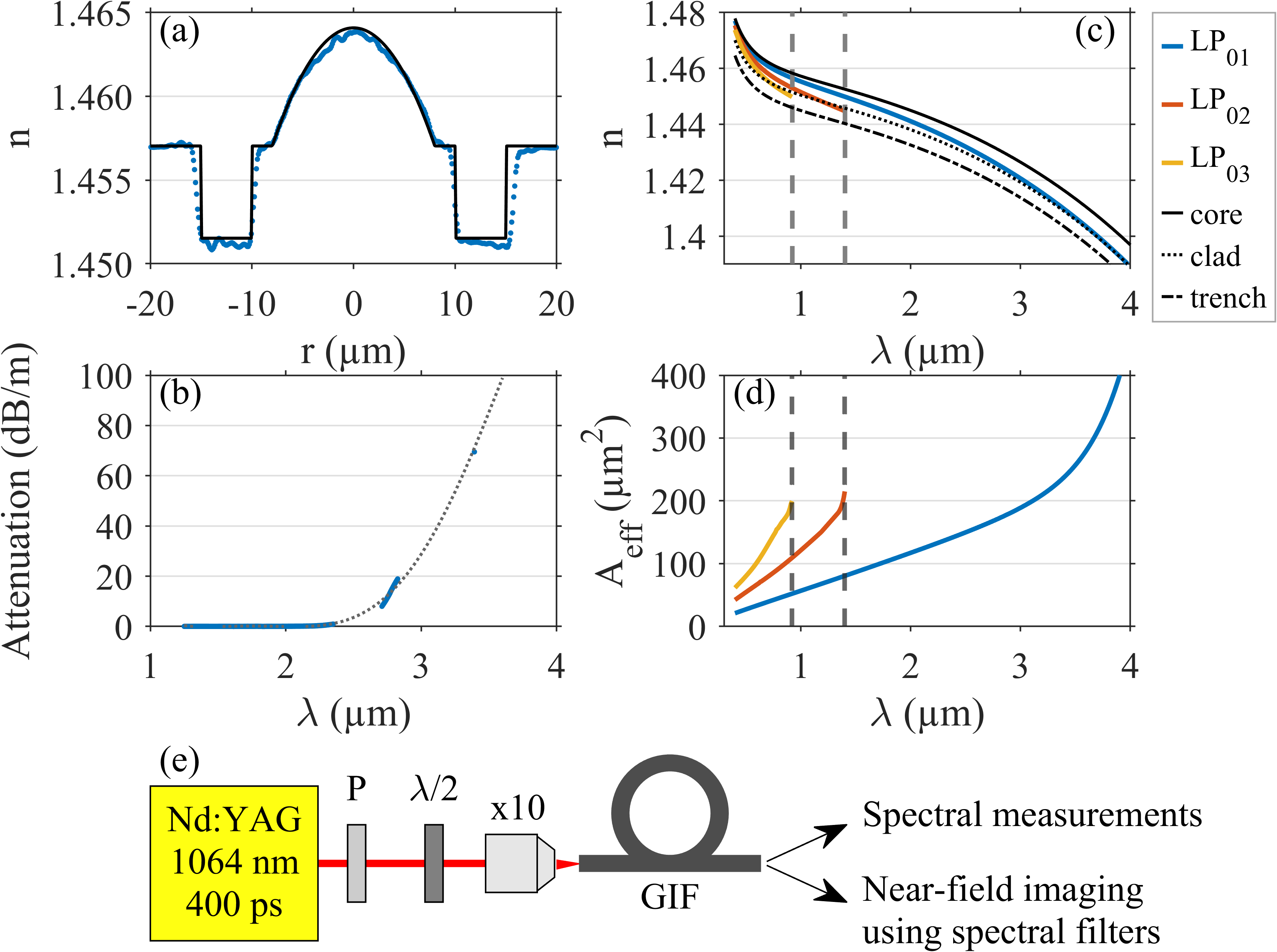}
    \caption{
    (a) Refractive index profile measured at \SI{0.632}{\micro\meter} (blue dots) and corresponding fit used for our numerical analysis (dotted line).
    (b) Fiber attenuation measured by means of the cutback technique (blue dots) and corresponding polynomial fit (black dotted line).
    (c) Refractive index and (d) effective mode area of the first three LP$_{0,m}$ modes with cutoff wavelength marked with gray dashed lines.
    (e) Experimental setup for observing far-detuned frequency conversion. P: polarizer; $\lambda/2$: half-wave plate; GIF: graded-index fiber.
    }
    \label{fig:setup}
    \end{figure*}
    \newpage
\section*{Experimental and numerical parameters}

We used a commercially-available few-mode GRIN fiber for our experiments. The refractive index profile of the fiber, measured at \SI{0.632}{\micro\meter} \cite{dupiol2017}, is shown in blue in Fig.~\ref{fig:setup}(a).
It can be accurately approximated (see black line in Fig.~\ref{fig:setup}(a)) by assuming the following opto-geometric parameters. We considered a fiber core with a radius of \SI{8}{\micro\meter} and a $\mathrm{GeO}_2$ doped parabolic profile, with a maximum concentration of \SI{5}{\percent\mol}. The core is then surrounded by a \SI{2}{\micro\meter}-thick layer of pure silica, followed by a \SI{5}{\micro\meter} fluorine-doped trench and further by the pure silica cladding. We assumed the refractive index shift induced by the fluorine to be \makebox{$ -5.5\cdot 10^{-3}$} (compared to pure silica).

The refractive index curves for the center of the germanium doped core, silica cladding, and fluorine trench as a function of wavelength are shown in Fig.~\ref{fig:setup}(c). We made use of Sunak material dispersion model for calculating the refractive index of both pure and GeO$_2$-doped silica \cite{sunak1989}. For such silica-based fibers, linear losses are typically negligible for meter-long segments within the visible and near-infrared spectral regions.
However, as the wavelength approaches the 2-µm waveband, the silica transparency significantly drops.
Consequently, we measured the fiber losses by means of the common cutback method combined with a home-made fiber supercontinuum source and a He-Ne laser emitting at \SI{3.39}{\micro\meter}.
The results are depicted in blue in Fig.~\ref{fig:setup}(b).
A polynomial fit is also reported with black dotted line to illustrate the full wavelength-dependent loss curve over the 1.25-4 µm spectral band. The attenuation reaches \SI{1}{dB/m} at \SI{2.4}{\micro\meter} and exceeds \SI{85}{dB/m} for wavelengths beyond \SI{3.5}{\micro\meter}.

Modal properties of the fiber were calculated using COMSOL Multiphysics software, only scalar linearly polarized LP$_{0,m}$ modes were considered. The effective refractive indices and effective mode areas of the first three LP$_{0,m}$ modes are plotted in Fig.~\ref{fig:setup}(c) and (d), respectively. LP$_{0,3}$ and LP$_{0,2}$ modes are cut off at 0.92 and \SI{1.40}{\micro\meter}, respectively.
Furthermore, the effective refractive index of the LP$_{0,1}$ mode approaches the cladding refractive index for wavelengths exceeding \SI{3.6}{\micro m}, thus indicating that the mode significantly propagates in the cladding, thus confirmed by the rise of its effective area in that spectral region.

To experimentally investigate intramodal and intermodal FWM processes in an optical fiber, we implemented the setup described just below, and depicted in Fig.~\ref{fig:setup}(e). 
We used a high-power laser source (Teem Photonics, PNP-M08010-1x0) operating at \SI{1.064}{\micro\meter} with a repetition rate of \SI{1}{kHz} that delivers \SI{400}{ps} pulses. The laser beam was passed through both polarizer and half-wave plate for power and polarization control, before being coupled into the fiber with a $\times10$ microscope objective.
Careful attention was paid to light coupling in such a way to maximize the excitation of the fundamental mode LP$_{0,1}$ of the fiber under test. We investigated only short fiber segments (less than 2 meters). The fiber output spectrum was monitored by using two optical spectrum analyzers (Yokogawa, AQ6374 and AQ6377), while the associated spatial content was analyzed using two beam profilers (Cinogy, CinCam CMOS-1202 IR; FLIR, SC7300) and a set of suitable spectral filters.

\section*{Results and discussion}

\begin{figure*}
    \centering\includegraphics[width=0.8\linewidth]{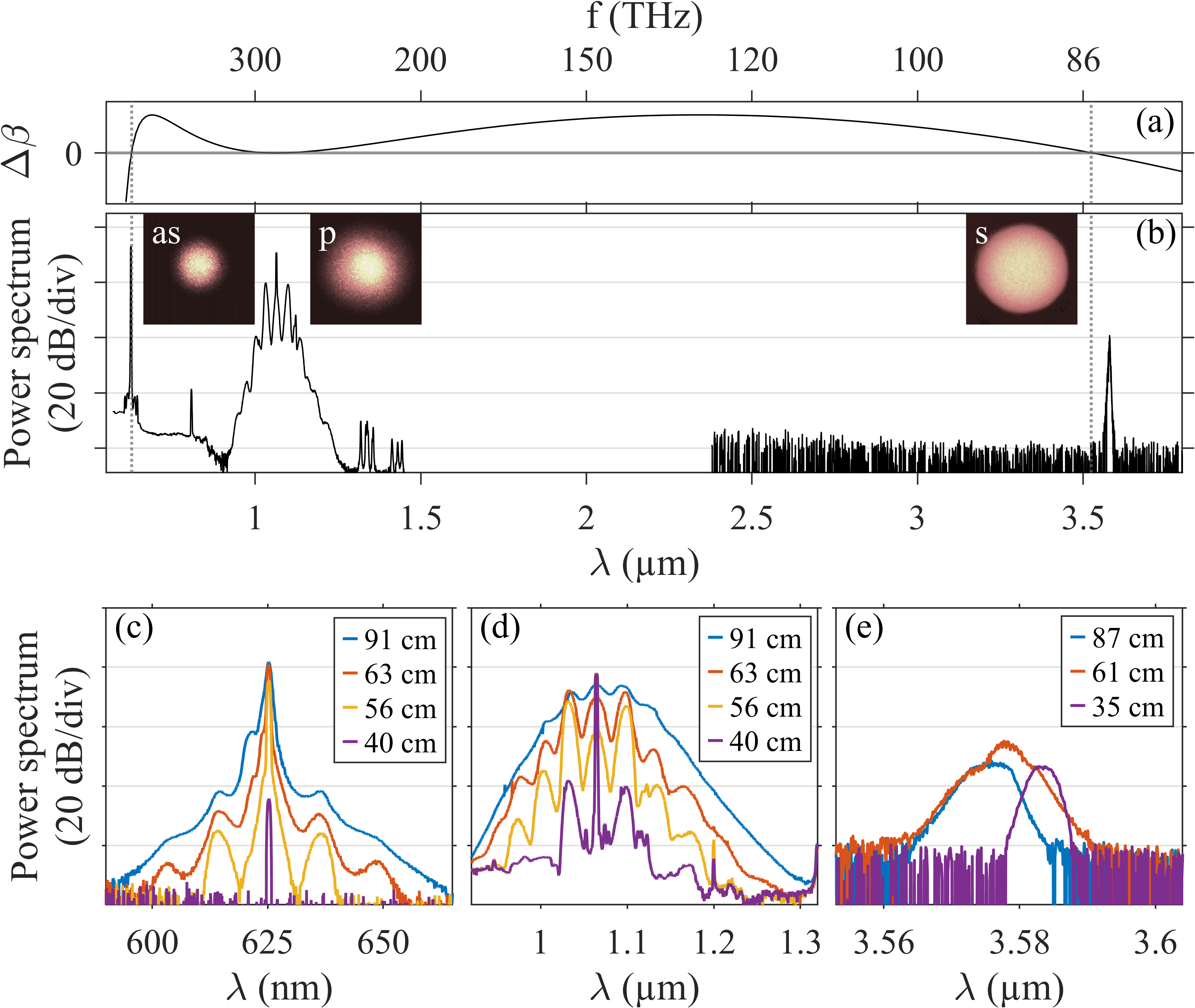}
    \caption{(a) Phase mismatch $\Delta\beta$ calculated for the FWM process in the fundamental mode. (b) Experimental spectrum recorded at the output of a \SI{70}{cm}-long fiber segment when the aforementioned FWM process in the fundamental mode is optimized. Insets: near-field images of the three spectral peaks (anti-Stokes, pump, and Stokes waves) recorded using spectral filters. (c-e) Corresponding spectra of anti-Stokes, pump, and Stokes waves respectively recorded at the output of fiber segments of increasing lengths.}
    \label{fig:lp01}
    \end{figure*}

In the following, we present and analyze two typical output spectra observed for slightly different average powers but similar spatial light coupling conditions into a \SI{70}{cm}-long fiber segment.
As a first example, we maximized the coupling into the fundamental mode LP$_{0,1}$ by monitoring the output modal content of the fiber. The corresponding output spectrum for an average power of \SI{27.5}{mW} (equivalent to a 60\% coupling factor of the input laser) is plotted in Fig.~\ref{fig:lp01}(b).
The pump spectrum is broadened around \SI{1.064}{\micro\meter} due to intermodal modulation instability arising from the bi-modal interaction with the higher-order mode LP$_{0,2}$~\cite{dupiol2017immi,stolen1975}. By finely analyzing the MI peak frequency and gain recorded for several fiber lengths (as indicated in the zoom of Fig.~\ref{fig:lp01}(d)), we can estimate the initial power distribution between the two spatial modes involved at the pump wavelength through the theoretical phase-matching~\cite{dupiol2017immi}. Our analysis reveals that more than 85$\%$ of the total power was injected into the fundamental mode. When looking at the overall spectrum shown in Fig.~\ref{fig:lp01}(b), we also checked that the residual peaks around 0.8 and \SI{1.4}{\micro\meter} arise from the laser source. However, the remarkable feature here is the emergence of two far-detuned peaks located at \SI{3.580}{\micro\meter} and \SI{0.625}{\micro\meter}, representing a symmetric detuning of approximately \SI{200}{THz} from the pump. These spectral positions correspond to the wavelengths of the Stokes and anti-Stokes of a FWM process involving solely the fundamental LP$_{0,1}$ mode. The near-field spatial characterization of filtered spectral peaks (see insets in Fig.~\ref{fig:lp01}(b)) confirms the intramodal nature of this FWM process. More specifically, we used a shortpass \SI{0.95}{\micro\meter} spectral filter to capture the anti-Stokes wave at \SI{0.625}{\micro\meter}, a bandpass \SI{1.064}{\micro\meter} spectral filter to capture the pump wave, and a longpass \SI{2}{\micro\meter} spectral filter to capture the Stokes wave at \SI{3.580}{\micro\meter}.
The spatial intensity patterns can all be attributed to the fundamental mode LP$_{0,1}$, whose full-width at half-maximum significantly broadens over the fiber core with increasing wavelength.
The power contained in the mid-infrared peak was measured to be \SI{40}{\micro\watt} when the total average power at the fiber output was \SI{27.5}{mW}. To the best of our knowledge, this observation is the most far-detuned frequency conversion into the mid-IR range performed in a silica-based fiber.

Any degenerate-pump-frequency (intra- or intermodal) FWM process occuring in an optical fiber can be described by the following linear phase-matching condition~\cite{nazemosadat2016}: 
$\Delta\beta=\beta^s_\mu + \beta^{as}_\eta - \beta^p_\nu - \beta^p_\kappa = 0$, where $\beta^p_\kappa$ represents the propagation constant of the spatial mode $\kappa$ at the pump frequency $\omega_p$. Superscripts $s,as$ represent the Stokes, and anti-Stokes frequencies, respectively. 
The above phase-matching condition can be analyzed as a function of the mode number~$m$, and we restrict our analysis to LP$_{0,m}$ modes. In practice, the propagation constant of mode LP$_{0,m}$ in a GRIN fiber can be simply calculated by the summation of the material dispersion and an approximate waveguide contribution~\cite{nazemosadat2016}. But this common picture of equidistant distribution of propagation constants fails to correctly describe our observations, nor the use of a finite-order Taylor-series expansion of propagation constants, without implying a misinterpretation of mode combinations involved in FWM processes. Consequently, the phase-matching curves presented here were obtained by taking into account the full frequency dependence of the propagation constants. As shown in Fig.~\ref{fig:lp01}(a), our results confirm that the phase-matching of the far-detuned frequency peaks observed at \SI{\pm200}{THz} can be fully driven by the higher-order dispersion of solely the fundamental mode $m=1$ of the GRIN fiber (i.e., a degenerate FWM that transfers energy from a strong pump wave to two waves). Indeed, the corresponding linear phase-matching accurately predicts the spectral positions of the Stokes and anti-Stokes at \SI{3.524}{\micro\meter} and \SI{0.627}{\micro\meter}, respectively. Note that the disagreement in frequency with experiments remains in the order of \SI{1}{THz} for both peaks.

\begin{figure*}
    \centering\includegraphics[width=0.85\linewidth]{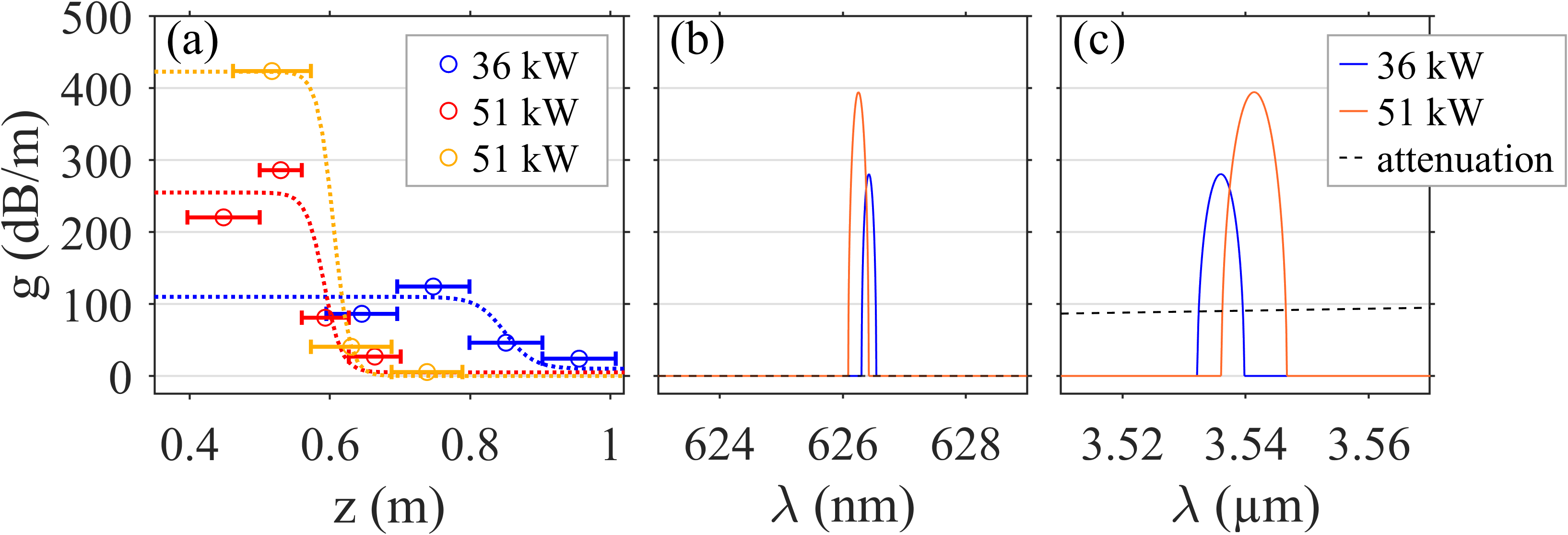}
    \caption{(a) Experimental gain measurements for the anti-Stokes wave centred at \SI{0.625}{\micro\meter} as a function of propagation distance $z$, obtained for peak powers in the fundamental mode equal to 51~kW (red and yellow) and 36~kW (blue) (with fitting curves in dashed lines as guides to the eye). Horizontal solid segments also indicate the cutback fiber lengths. (b and c) Theoretical gain calculations for the anti-Stokes (b) and Stokes (c) waves for 36~kW (blue) and 51~kW (orange) peak power. Black dashed line indicates the value of fiber attenuation in this spectral region.}
    \label{fig:gain}
    \end{figure*}

To further study this FWM process, we compare the power spectra of anti-Stokes and pump waves at different propagation distances with a cutback experiment.  We plot some selected spectra of anti-Stokes in Fig.~\ref{fig:lp01}(c), obtained for different fiber lengths indicated in the legend, with input beam launch conditions kept unchanged (except average power here equals to \SI{24}{mW}). Notably, the anti-Stokes power grows significantly with propagation distance. After approximately \SI{70}{cm}, the spontaneous gain for the anti-Stokes saturates and further propagation leads solely to spectral broadening. The estimation of this growth rate along the propagation distance is analyzed later (see Fig.~\ref{fig:gain}(a)). The anti-Stokes broadening is induced by the cascaded broadening of the pump spectrum, as observed in Fig.~\ref{fig:lp01}(d), through cross-phase modulation effects. Note that the same frequency detuning of spectral sidebands is observed (approximately $8.5$~THz).
Additionally, in Fig.~\ref{fig:lp01}(e), we present some selected spectra of the mid-IR Stokes wave obtained for three different fiber lengths. After each cutback the fiber alignment was optimized to achieve the highest intensity of the mid-IR peak (for a constant average power of \SI{27.5}{mW}). Clearly, the Stokes power is limited by the strong fiber losses in this spectral region (see Fig.~\ref{fig:setup}(b)). After exceeding approximately \SI{70}{cm} of propagation, the Stokes power begins to drop as fiber losses are overcoming the already saturated gain of the FWM process. The changes observed in the Stokes wavelength and bandwidth in Fig.~\ref{fig:lp01}(e) can be ascribed to the slight input coupling variations, but also to the pump depletion and spectral broadening occurring along the fiber length.

    \begin{figure*}
    \centering\includegraphics[width=0.8\linewidth]{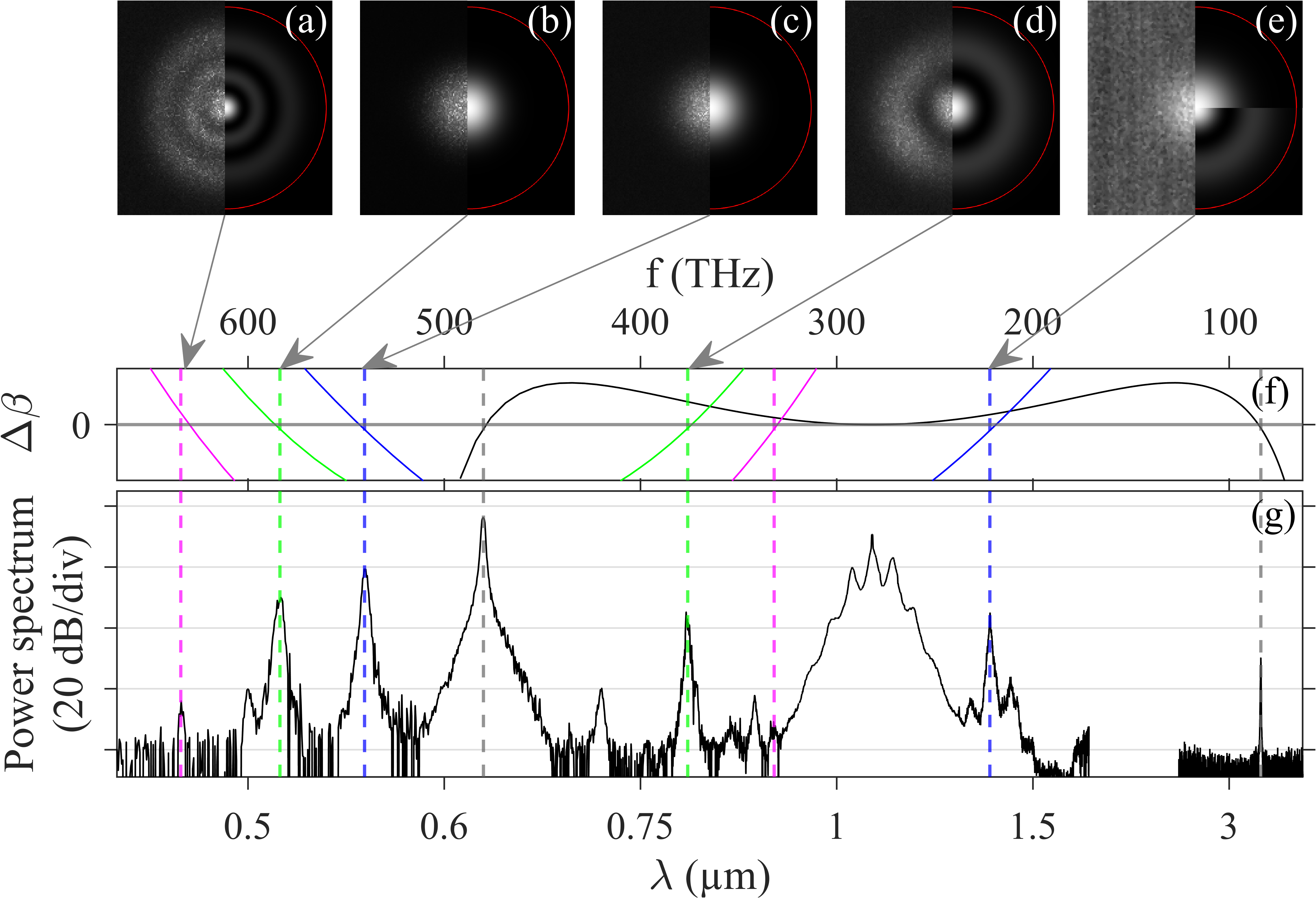}
    \caption{(a-e) Near-field images of the spectrally filtered peaks indicated by arrows and colored dashed lines in subplot (g). Left sections are experimental measurements, whereas right sections provide corresponding calculated mode distributions. In subplot (e), both calculated LP$_{0,1}$ and LP$_{0,2}$ modes are shown. Red line indicates the inner edge of the fluorine-doped trench.
    (f) Phase mismatch $\Delta\beta$ calculated for intramodal and intermodal FWM processes (details provided in text). (g) Experimental spectrum recorded at the output of a \SI{70}{cm}-long fiber segment with an average power of \SI{29.8}{mW}.}
    \label{fig:cascade}
    \end{figure*}

As another example, we increased the input peak power into our 70-cm-long GRIN fiber by slightly adjusting the position of the fiber around the input focal plane, following which we collected the output spectrum shown in Fig.~\ref{fig:cascade}(g). The average power at the fiber output is now \SI{29.8}{mW}, thus corresponding to a coupling efficiency of 65\%. Here, we obtained a similar spectrum characterized by a broadened pump spectrum around \SI{1.064}{\micro\meter} and two far-detuned frequency peaks at $\pm$200~THz. However, a new set of frequency peaks has emerged in the visible region, as previously reported in Ref.~\cite{dupiol2017} in a slightly longer fiber segment (1~meter). The minor adjustment in the input conditions modified the characteristic frequency of intermodal MI bands around the pump (now about 10 THz detuning). Consequently, we estimate that no more than 80$\%$ of the total power is now injected into the fundamental mode (and the remaining into the LP$_{0,2}$ mode). However, as previously shown, the fundamental mode remains the main mode involved in the distinct FWM processes studied below. A detailed investigation of each phase-matched FWM at the origin of the spectral peaks was performed both experimentally and theoretically. The spatial content for the newly emerged spectral peaks was characterized by near-field imaging as shown in Figs~\ref{fig:cascade}(a-e). Various configurations of intermodal FWM processes were calculated and analyzed in Table~\ref{tab:1}. It appears that the superposition of both intramodal and intermodal FWM processes could emerge, as very similar phase-matched frequencies can be found. Consequently, beyond the necessary condition of phase-matching for the FWM interactions for distinct mode combinations, the relative strength of the interaction has to be determined by
the nonlinear coupling coefficient $\gamma_{1,2,3,4} = \frac{n_2\omega_1}{c}f_{1,2,3,4}$ driven by the overlap integral of the field distributions 
$f_{1,2,3,4} = \left|\frac{ \langle F_1^* F_2^* F_3 F_4\rangle }
            { \sqrt{\langle\left|F_1\right|^2\rangle
                    \langle\left|F_2\right|^2\rangle
                    \langle\left|F_3\right|^2\rangle
                    \langle\left|F_4\right|^2\rangle }  }\right| $
~\cite{agrawalbook}, where angle brackets denote integration over the transverse coordinates and subscripts indicate four interacting waves. $F_j$ is the spatial distribution of the fiber mode in which the $j$th field propagates inside the fiber.
It is noteworthy to emphasize that for the far-detuned process, the integration involves the wavelength-dependent field distributions.
For the case of the far-detuned peaks at $\pm$200~THz previously studied and observed again, there are two possible FWM processes involving combinations of LP$_{0,1}$ mode only, or with LP$_{0,2}$ modes in Table~\ref{tab:1}, namely: $\{\mu,\eta,\nu,\kappa\} = \{1,1,1,1\}$, or $\{1,2,1,2\}$.
These combinations exhibit different overlap integrals (which translates into the FWM efficiencies), thus confirming that this process under-study predominantly involves the fundamental mode (the corresponding $\Delta\beta$ is depicted with black line in Fig.~\ref{fig:cascade}(f)). In this regard we also point out that the resulting FWM efficiency also depends on the power contained in the pumping modes $\{\nu,\kappa\}$.

The second pair of analyzed peaks is located at \SI{0.555}{\micro\meter} and \SI{1.352}{\micro\meter} (blue dashed lines in Fig.~\ref{fig:cascade}(g)). In that case, we retrieve experimentally the main signature of the fundamental mode for such peaks, see Figs.~\ref{fig:cascade}(c) and (e). Table~\ref{tab:1} provides the corresponding FWM processes with the highest overlap integrals. The origin of the observed spectral components can only be explained by involving non-degenerate FWM with two distinct pump wavelengths, specifically \SI{0.625}{\micro\meter} and \SI{1.064}{\micro\meter}, for satisfying the condition $\omega_s + \omega_{as} = \omega_{p1} + \omega_{p2}$. The mode combination $\{2,1,1,1\}$ appears as the most probable scenario consistent with the pump wavelengths mostly carried by the fundamental mode (the corresponding $\Delta\beta$ is depicted with blue line in Fig.~\ref{fig:cascade}(f)). However, one has to point out that the LP$_{0,2}$ mode is close to cutoff at the Stokes wavelength, thus explaining why the Stokes peak is observed at \SI{1.352}{\micro\meter} in the fundamental mode with a lower intensity than the anti-Stokes wave at \SI{0.555}{\micro\meter}. Other combinations listed in Table~\ref{tab:1} provide FWM processes with lower efficiency but enabling Stokes emission in the LP$_{0,1}$ mode with mixed pumping in both LP$_{0,1}$ and LP$_{0,2}$ modes.
The third pair of peaks emerges at \SI{0.514}{\micro\meter} and \SI{0.798}{\micro\meter} (green dashed lines in Fig.~\ref{fig:cascade}(g)). These sidebands are mainly associated with the LP$_{0,1}$ and LP$_{0,2}$ modes, respectively, see Figs.~\ref{fig:cascade}(b) and (d). When checking possible phase-matched combinations, a degenerate FWM with pumping in the fundamental mode at \SI{0.625}{\micro\meter} was identified as the process with the highest overlap integral, thus giving $\{2,1,1,1\}$ (the corresponding $\Delta\beta$ is depicted with green line in Fig.~\ref{fig:cascade}(f)). This is in excellent agreement with our above measurements. Other combinations are also listed in Table~\ref{tab:1} for comparison.
Finally, the fourth pair of peaks recorded at \SI{0.473}{\micro\meter} and \SI{0.904}{\micro\meter} (magenta dashed lines in Fig.~\ref{fig:cascade}(g)) exhibits low intensities, only the spatial content of anti-Stokes wave was recorded, and it is mainly supported by the LP$_{0,3}$ mode, see Fig.~\ref{fig:cascade}(a). Our analysis indicates that these wavelengths can be associated with degenerate FWM processes with pumping at \SI{0.625}{\micro\meter}, and involving the LP$_{0,1}$ and LP$_{0,2}$ modes. The one corresponding to our experimental characterization refers to
$\{2,3,1,2\}$ in Table~\ref{tab:1}, and confirms the low efficiency (the corresponding $\Delta\beta$ is depicted with magenta line in Fig.~\ref{fig:cascade}(f)).

\begin{table}
\centering
\caption{Comparison of calculated FWM combinations with the highest overlap integrals (up to three cases for each process). For comparison, the self-phase modulation coefficient at pump wavelength in the fundamental mode is \SI{0.0165}{\micro\meter^{-2}}.}
\begin{tabular}{c||c|c|c|c||c}

    $\{\mu,\eta,\nu,\kappa\}$ &
    $\lambda^{s}_\mu $ &
    $\lambda^{as}_\eta $ &
    $\lambda^{p1}_\nu $ &
    $\lambda^{p2}_\kappa $ &
    $f_{s,as,p1,p2} \left(\upmu\mathrm{m}^{-2}\right)$ \\\hline
    $\{1, 1, 1, 1\}$ & 3.524 & 0.627 & 1.064 & 1.064 & 0.0111 \\
    $\{1, 2, 1, 2\}$ & 3.475 & 0.629 & 1.064 & 1.064 & 0.0045 \\\hline
    
    $\{2, 1, 1, 1\}$ & 1.351 & 0.556 & 1.064 & 0.625 & 0.0135 \\
    $\{2, 2, 2, 1\}$ & 1.364 & 0.554 & 1.064 & 0.625 & 0.0059 \\
    $\{1, 3, 1, 2\}$ & 1.373 & 0.552 & 1.064 & 0.625 & 0.0054 \\\hline
    
    $\{2, 1, 1, 1\}$ & 0.799 & 0.513 & 0.625 & 0.625 & 0.0171 \\
    $\{3, 1, 1, 2\}$ & 0.791 & 0.517 & 0.625 & 0.625 & 0.0111 \\
    $\{1, 2, 1, 1\}$ & 0.802 & 0.512 & 0.625 & 0.625 & 0.0106 \\\hline
    
    $\{2, 2, 1, 1\}$ & 0.906 & 0.477 & 0.625 & 0.625 & 0.0119 \\
    $\{3, 1, 1, 1\}$ & 0.894 & 0.480 & 0.625 & 0.625 & 0.0113 \\
    $\{2, 3, 1, 2\}$ & 0.909 & 0.476 & 0.625 & 0.625 & 0.0073 \\
\end{tabular}
\label{tab:1}
\end{table}

We clearly demonstrate that the additional FWM processes observed in the second experiment find their origin in the combination of the secondary pump at \SI{0.625}{\micro\meter} with or without the initial pump at \SI{1.064}{\micro\meter} in the fundamental mode. This phenomenon simply arises from the significant power contained by the fundamental mode at those wavelengths. More specifically, the secondary pump is enabled by the high gain provided by the first intramodal degenerate FWM process studied in Fig.~\ref{fig:lp01}(b). To go beyond the analysis of phase-matching, we characterize the gain of this spontaneous far-detuned FWM process due to a strong pump wave in the LP$_{0,1}$ mode.

By following the general derivation of the parametric FWM gain from Ref.~\cite{agrawalbook}, we can limit the calculation to a set of three coupled nonlinear equations since the pump fields cannot be distinguished on the basis of their frequencies, polarization, or spatial profile in the present case. Our analysis assumes an undepleted pump but it includes the frequency-dependence of overlap integrals. We find that the corresponding effective phase-mismatch becomes $\Delta k = \Delta\beta - 2\gamma_{p,p}P_0 + 2\gamma_{s,p}P_0 + 2\gamma_{as,p}P_0$, where $\Delta\beta =\beta^s_1 + \beta^{as}_1 - 2\beta^p_1$ is the phase-matching condition in the linear regime; $\gamma_{p,p}$,  $\gamma_{s,p}$ and $\gamma_{as,p}$ denote the effective nonlinearity coefficients for pump's self-phase modulation (SPM), pump-induced cross-phase modulation (XPM) of Stokes and anti-Stokes bands, respectively; and $P_0$ is the input peak power (here: in the fundamental mode, 85$\%$ of \SI{60}{kW}). The gain $g$ depends on the pump power and is defined as $g=\sqrt{(\gamma_\mathrm{FWM}P_0)^2-(\Delta k / 2)^2}$.
The maximum gain ($g_\mathrm{max}$) is then achieved for $\Delta k = 0$, thus $g_\mathrm{max} = \gamma_\mathrm{FWM}P_0$, with $\gamma_\mathrm{FWM} = \sqrt{\gamma_{s,as,p,p}\gamma_{as,s,p,p}}$ being the effective nonlinearity for the intramodal FWM process.
Figure~\ref{fig:gain} compares some experimental gain measurements along the fiber length (Fig.~\ref{fig:gain}(a)) with theoretical predictions for two peak powers (Fig.~\ref{fig:gain}(b-c)).
Estimation of experimental gain was possible only for the anti-Stokes wave in the visible region, based on several recordings of spectra as shown in Fig.~\ref{fig:lp01}(c).
Such a gain inference over very short propagation distances is tricky. As shown by our two data sets for the same power of \SI{51}{kW} in the fundamental mode, the gain estimation strongly depends on the cutback increment (indicated by horizontal segments in Fig.~\ref{fig:gain}(a)). Nevertheless, our gain measurements agree well with our theoretical predictions close to \SI{400} dB/m. Our measurement also shows the complete saturation of the FWM process after \SI{0.7}{\meter} as the gain rapidly drops to zero. It is also worth to mention that our theoretical phase-matching $\Delta k$ (including contributions of SPM and XPMs) better predicts the spectral positions of the Stokes and anti-Stokes at \SI{3.545}{\micro\meter} and \SI{0.626}{\micro\meter}, respectively.

For $P_0=$ \SI{36}{kW}, our gain estimate is lower than theory, this can be also attributed to the fact that gain saturation occurs later along the fiber due to lower input power.
Consequently, several effects can impact gain efficiency of this FWM process, namely the mid-IR higher losses accumulated over propagation, the competition with other FWM processes such as intermodal MI around the pump, and the temporal walk-off between the generated peaks and the pump \cite{agrawalbook}.
In the latter case, the group velocity mismatch is \SI{48}{ps/m} between the pump and anti-Stokes wavelengths, and \SI{168}{ps/m} between the pump and Stokes wavelengths.
As a result, one can calculate the corresponding walk-off lengths $L_W$ (for a characteristic pulse duration $T_0=T_{FWHM}/1.665=$ \SI{240}{ps}), we found $L_W=$ \SI{5} and \SI{1.4}{m}, respectively, which is longer than our fiber segments under study.
A clear identification of the involved detrimental effects would require further numerical studies based on the complex modeling of MMF as a function of modal excitation \cite{bejot2019multimodal}.

Here it is worth highlighting that an extreme value of parametric gain (\SI{400} dB/m) was obtained, even for far-detuned frequencies.
Typically, in single-mode and small-core fibers, far-detuned frequency conversion in the normal dispersion regime has already been investigated, but rather limited to frequency shifts of tens of THz and FWM gain of tens of dB/m \cite{harvey2003,godin2014}, without taking into account the frequency-dependence of overlap integrals.
Here extended FWM features are supported by the combination of the strong mode confinement over a large spectral bandwidth linked to GRIN profile and high peak powers. In Fig.~\ref{fig:setup}(d), one can notice that the effective area of the fundamental mode remains in the 10 to \SI{200}{\micro\meter^2} range over the \SI{3}{\micro\meter} spectral bandwidth. The linear behavior then drastically changes around \SI{3.5}{\micro\meter} and can be also associated with higher confinement losses (possibly explaining the low power measured for Stokes peak in the mid-IR).  
Such far-detuned phase-matchings are also possible theoretically in standard step-index profiles of silica fibers. We numerically considered the case of step-index fiber with a \SI{5}{\percent\mol} $\mathrm{GeO}_2$-doped core with a radius of \SI{8}{\micro\meter}. The analysis of phase-matching predicts the spectral position of Stokes and anti-Stokes peaks at \SI{3.411}{\micro\meter} and \SI{0.630}{\micro\meter}, respectively. However the calculated overlap integral equal to \SI{0.0063}{\micro\meter^{-2}} is almost twice as low as the case of the GRIN fiber studied here. The lower modal confinement (i.e., effective mode area varies from 100 to \SI{300}{\micro\meter^2}) does not result in such high FWM gain. Experimentally, we also checked that our available few-mode step-index silica fibers at \SI{1.064}{\micro\meter} did not allow us to observe any far-detuned FWM peaks with similar pumping conditions.

\section*{Conclusion}

In summary, we have reported a far-detuned wavelength conversion process with high gain in a few-mode GRIN fiber that extends into the mid-infrared well-beyond the common transparency window of silica. Extreme features of parametric conversion were obtained by means of a commercially-available GRIN profile, though there is still room for optimization procedure in the fiber design to tailor phase-matched frequencies and parametric gain in many spectral windows of interest. Moreover, our results highlight that broadband wavelength conversion process into MMFs still deserves further detailed investigations, especially when considering the fine spatial-spectral structure \cite{leventoux2021}.

\begin{acknowledgments}

\section*{Funding}
The authors acknowledge funding from French Ministry for Europe and Foreign Affairs, French Ministry for Higher Education and Research (through French-Polish Polonium Hubert Curien Partnership), Polish National Agency for Academic Exchange (PHC Polonium), and National Science Center, Poland (2018/30/E/ST7/00862). K.S. acknowledges support by Bourse du Gouvernement Français from French Government and by STER Programme Internationalisation of Wroclaw University of Science and Technology Doctoral School from Polish National Agency for Academic Exchange.

\end{acknowledgments}

\bibliography{library}

\begin{thebibliography}{23}%
\makeatletter
\providecommand \@ifxundefined [1]{%
 \@ifx{#1\undefined}
}%
\providecommand \@ifnum [1]{%
 \ifnum #1\expandafter \@firstoftwo
 \else \expandafter \@secondoftwo
 \fi
}%
\providecommand \@ifx [1]{%
 \ifx #1\expandafter \@firstoftwo
 \else \expandafter \@secondoftwo
 \fi
}%
\providecommand \natexlab [1]{#1}%
\providecommand \enquote  [1]{``#1''}%
\providecommand \bibnamefont  [1]{#1}%
\providecommand \bibfnamefont [1]{#1}%
\providecommand \citenamefont [1]{#1}%
\providecommand \href@noop [0]{\@secondoftwo}%
\providecommand \href [0]{\begingroup \@sanitize@url \@href}%
\providecommand \@href[1]{\@@startlink{#1}\@@href}%
\providecommand \@@href[1]{\endgroup#1\@@endlink}%
\providecommand \@sanitize@url [0]{\catcode `\\12\catcode `\$12\catcode `\&12\catcode `\#12\catcode `\^12\catcode `\_12\catcode `\%12\relax}%
\providecommand \@@startlink[1]{}%
\providecommand \@@endlink[0]{}%
\providecommand \url  [0]{\begingroup\@sanitize@url \@url }%
\providecommand \@url [1]{\endgroup\@href {#1}{\urlprefix }}%
\providecommand \urlprefix  [0]{URL }%
\providecommand \Eprint [0]{\href }%
\providecommand \doibase [0]{https://doi.org/}%
\providecommand \selectlanguage [0]{\@gobble}%
\providecommand \bibinfo  [0]{\@secondoftwo}%
\providecommand \bibfield  [0]{\@secondoftwo}%
\providecommand \translation [1]{[#1]}%
\providecommand \BibitemOpen [0]{}%
\providecommand \bibitemStop [0]{}%
\providecommand \bibitemNoStop [0]{.\EOS\space}%
\providecommand \EOS [0]{\spacefactor3000\relax}%
\providecommand \BibitemShut  [1]{\csname bibitem#1\endcsname}%
\let\auto@bib@innerbib\@empty
\bibitem [{\citenamefont {Wright}\ \emph {et~al.}(2015)\citenamefont {Wright}, \citenamefont {Christodoulides},\ and\ \citenamefont {Wise}}]{wright2015}%
  \BibitemOpen
  \bibfield  {author} {\bibinfo {author} {\bibfnamefont {L.~G.}\ \bibnamefont {Wright}}, \bibinfo {author} {\bibfnamefont {D.~N.}\ \bibnamefont {Christodoulides}},\ and\ \bibinfo {author} {\bibfnamefont {F.~W.}\ \bibnamefont {Wise}},\ }\bibfield  {title} {\bibinfo {title} {Controllable spatiotemporal nonlinear effects in multimode fibres},\ }\href {https://doi.org/10.1038/nphoton.2015.61} {\bibfield  {journal} {\bibinfo  {journal} {Nature Photonics}\ }\textbf {\bibinfo {volume} {9}},\ \bibinfo {pages} {306} (\bibinfo {year} {2015})}\BibitemShut {NoStop}%
\bibitem [{\citenamefont {Krupa}\ \emph {et~al.}(2016{\natexlab{a}})\citenamefont {Krupa}, \citenamefont {Tonello}, \citenamefont {Barth{\'e}l{\'e}my}, \citenamefont {Couderc}, \citenamefont {Shalaby}, \citenamefont {Bendahmane}, \citenamefont {Millot},\ and\ \citenamefont {Wabnitz}}]{krupa2016}%
  \BibitemOpen
  \bibfield  {author} {\bibinfo {author} {\bibfnamefont {K.}~\bibnamefont {Krupa}}, \bibinfo {author} {\bibfnamefont {A.}~\bibnamefont {Tonello}}, \bibinfo {author} {\bibfnamefont {A.}~\bibnamefont {Barth{\'e}l{\'e}my}}, \bibinfo {author} {\bibfnamefont {V.}~\bibnamefont {Couderc}}, \bibinfo {author} {\bibfnamefont {B.~M.}\ \bibnamefont {Shalaby}}, \bibinfo {author} {\bibfnamefont {A.}~\bibnamefont {Bendahmane}}, \bibinfo {author} {\bibfnamefont {G.}~\bibnamefont {Millot}},\ and\ \bibinfo {author} {\bibfnamefont {S.}~\bibnamefont {Wabnitz}},\ }\bibfield  {title} {\bibinfo {title} {Observation of geometric parametric instability induced by the periodic spatial self-imaging of multimode waves},\ }\href {https://doi.org/10.1103/PhysRevLett.116.183901} {\bibfield  {journal} {\bibinfo  {journal} {Physical Review Letters}\ }\textbf {\bibinfo {volume} {116}},\ \bibinfo {pages} {183901} (\bibinfo {year} {2016}{\natexlab{a}})}\BibitemShut {NoStop}%
\bibitem [{\citenamefont {Krupa}\ \emph {et~al.}(2017)\citenamefont {Krupa}, \citenamefont {Tonello}, \citenamefont {Shalaby}, \citenamefont {Fabert}, \citenamefont {Barth{\'e}l{\'e}my}, \citenamefont {Millot}, \citenamefont {Wabnitz},\ and\ \citenamefont {Couderc}}]{krupa2017}%
  \BibitemOpen
  \bibfield  {author} {\bibinfo {author} {\bibfnamefont {K.}~\bibnamefont {Krupa}}, \bibinfo {author} {\bibfnamefont {A.}~\bibnamefont {Tonello}}, \bibinfo {author} {\bibfnamefont {B.~M.}\ \bibnamefont {Shalaby}}, \bibinfo {author} {\bibfnamefont {M.}~\bibnamefont {Fabert}}, \bibinfo {author} {\bibfnamefont {A.}~\bibnamefont {Barth{\'e}l{\'e}my}}, \bibinfo {author} {\bibfnamefont {G.}~\bibnamefont {Millot}}, \bibinfo {author} {\bibfnamefont {S.}~\bibnamefont {Wabnitz}},\ and\ \bibinfo {author} {\bibfnamefont {V.}~\bibnamefont {Couderc}},\ }\bibfield  {title} {\bibinfo {title} {Spatial beam self-cleaning in multimode fibres},\ }\href {https://doi.org/10.1038/nphoton.2017.32} {\bibfield  {journal} {\bibinfo  {journal} {Nature Photonics}\ }\textbf {\bibinfo {volume} {11}},\ \bibinfo {pages} {237} (\bibinfo {year} {2017})}\BibitemShut {NoStop}%
\bibitem [{\citenamefont {Renninger}\ and\ \citenamefont {Wise}(2013)}]{renninger2013}%
  \BibitemOpen
  \bibfield  {author} {\bibinfo {author} {\bibfnamefont {W.~H.}\ \bibnamefont {Renninger}}\ and\ \bibinfo {author} {\bibfnamefont {F.~W.}\ \bibnamefont {Wise}},\ }\bibfield  {title} {\bibinfo {title} {Optical solitons in graded-index multimode fibres},\ }\href {https://doi.org/10.1038/ncomms2739} {\bibfield  {journal} {\bibinfo  {journal} {Nature Communications}\ }\textbf {\bibinfo {volume} {4}},\ \bibinfo {pages} {1719} (\bibinfo {year} {2013})}\BibitemShut {NoStop}%
\bibitem [{\citenamefont {Stefa{\'n}ska}\ \emph {et~al.}(2023)\citenamefont {Stefa{\'n}ska}, \citenamefont {B{\'e}jot}, \citenamefont {Tarnowski},\ and\ \citenamefont {Kibler}}]{stef2023}%
  \BibitemOpen
  \bibfield  {author} {\bibinfo {author} {\bibfnamefont {K.}~\bibnamefont {Stefa{\'n}ska}}, \bibinfo {author} {\bibfnamefont {P.}~\bibnamefont {B{\'e}jot}}, \bibinfo {author} {\bibfnamefont {K.}~\bibnamefont {Tarnowski}},\ and\ \bibinfo {author} {\bibfnamefont {B.}~\bibnamefont {Kibler}},\ }\bibfield  {title} {\bibinfo {title} {Experimental observation of the spontaneous emission of a space-time wavepacket in a multimode optical fiber},\ }\href {https://doi.org/10.1021/acsphotonics.2c01863} {\bibfield  {journal} {\bibinfo  {journal} {ACS Photonics}\ }\textbf {\bibinfo {volume} {10}},\ \bibinfo {pages} {727} (\bibinfo {year} {2023})}\BibitemShut {NoStop}%
\bibitem [{\citenamefont {Nazemosadat}\ \emph {et~al.}(2016)\citenamefont {Nazemosadat}, \citenamefont {Pourbeyram},\ and\ \citenamefont {Mafi}}]{nazemosadat2016}%
  \BibitemOpen
  \bibfield  {author} {\bibinfo {author} {\bibfnamefont {E.}~\bibnamefont {Nazemosadat}}, \bibinfo {author} {\bibfnamefont {H.}~\bibnamefont {Pourbeyram}},\ and\ \bibinfo {author} {\bibfnamefont {A.}~\bibnamefont {Mafi}},\ }\bibfield  {title} {\bibinfo {title} {Phase matching for spontaneous frequency conversion via four-wave mixing in graded-index multimode optical fibers},\ }\href {https://doi.org/10.1364/JOSAB.33.000144} {\bibfield  {journal} {\bibinfo  {journal} {J. Opt. Soc. Am. B}\ }\textbf {\bibinfo {volume} {33}},\ \bibinfo {pages} {144} (\bibinfo {year} {2016})}\BibitemShut {NoStop}%
\bibitem [{\citenamefont {Perret}\ \emph {et~al.}(2019)\citenamefont {Perret}, \citenamefont {Fanjoux}, \citenamefont {Bigot}, \citenamefont {Fatome}, \citenamefont {Millot}, \citenamefont {Dudley},\ and\ \citenamefont {Sylvestre}}]{perrey2019}%
  \BibitemOpen
  \bibfield  {author} {\bibinfo {author} {\bibfnamefont {S.}~\bibnamefont {Perret}}, \bibinfo {author} {\bibfnamefont {G.}~\bibnamefont {Fanjoux}}, \bibinfo {author} {\bibfnamefont {L.}~\bibnamefont {Bigot}}, \bibinfo {author} {\bibfnamefont {J.}~\bibnamefont {Fatome}}, \bibinfo {author} {\bibfnamefont {G.}~\bibnamefont {Millot}}, \bibinfo {author} {\bibfnamefont {J.~M.}\ \bibnamefont {Dudley}},\ and\ \bibinfo {author} {\bibfnamefont {T.}~\bibnamefont {Sylvestre}},\ }\bibfield  {title} {\bibinfo {title} {Supercontinuum generation by intermodal four-wave mixing in a step-index few-mode fibre},\ }\href {https://doi.org/10.1063/1.5045645} {\bibfield  {journal} {\bibinfo  {journal} {APL Photonics}\ }\textbf {\bibinfo {volume} {4}},\ \bibinfo {pages} {022905} (\bibinfo {year} {2019})}\BibitemShut {NoStop}%
\bibitem [{\citenamefont {Krupa}\ \emph {et~al.}(2016{\natexlab{b}})\citenamefont {Krupa}, \citenamefont {Louot}, \citenamefont {Couderc}, \citenamefont {Fabert}, \citenamefont {Guenard}, \citenamefont {Shalaby}, \citenamefont {Tonello}, \citenamefont {Pagnoux}, \citenamefont {Leproux}, \citenamefont {Bendahmane}, \citenamefont {Dupiol}, \citenamefont {Millot},\ and\ \citenamefont {Wabnitz}}]{krupa2016spatiotemporal}%
  \BibitemOpen
  \bibfield  {author} {\bibinfo {author} {\bibfnamefont {K.}~\bibnamefont {Krupa}}, \bibinfo {author} {\bibfnamefont {C.}~\bibnamefont {Louot}}, \bibinfo {author} {\bibfnamefont {V.}~\bibnamefont {Couderc}}, \bibinfo {author} {\bibfnamefont {M.}~\bibnamefont {Fabert}}, \bibinfo {author} {\bibfnamefont {R.}~\bibnamefont {Guenard}}, \bibinfo {author} {\bibfnamefont {B.~M.}\ \bibnamefont {Shalaby}}, \bibinfo {author} {\bibfnamefont {A.}~\bibnamefont {Tonello}}, \bibinfo {author} {\bibfnamefont {D.}~\bibnamefont {Pagnoux}}, \bibinfo {author} {\bibfnamefont {P.}~\bibnamefont {Leproux}}, \bibinfo {author} {\bibfnamefont {A.}~\bibnamefont {Bendahmane}}, \bibinfo {author} {\bibfnamefont {R.}~\bibnamefont {Dupiol}}, \bibinfo {author} {\bibfnamefont {G.}~\bibnamefont {Millot}},\ and\ \bibinfo {author} {\bibfnamefont {S.}~\bibnamefont {Wabnitz}},\ }\bibfield  {title} {\bibinfo {title} {Spatiotemporal characterization of supercontinuum extending from the visible to the mid-infrared in a multimode graded-index optical
  fiber},\ }\href {https://doi.org/10.1364/OL.41.005785} {\bibfield  {journal} {\bibinfo  {journal} {Opt. Lett.}\ }\textbf {\bibinfo {volume} {41}},\ \bibinfo {pages} {5785} (\bibinfo {year} {2016}{\natexlab{b}})}\BibitemShut {NoStop}%
\bibitem [{\citenamefont {Lopez-Aviles}\ \emph {et~al.}(2019)\citenamefont {Lopez-Aviles}, \citenamefont {Wu}, \citenamefont {Eznaveh}, \citenamefont {Eftekhar}, \citenamefont {Wise}, \citenamefont {Correa},\ and\ \citenamefont {Christodoulides}}]{lopez2019}%
  \BibitemOpen
  \bibfield  {author} {\bibinfo {author} {\bibfnamefont {H.~E.}\ \bibnamefont {Lopez-Aviles}}, \bibinfo {author} {\bibfnamefont {F.~O.}\ \bibnamefont {Wu}}, \bibinfo {author} {\bibfnamefont {Z.~S.}\ \bibnamefont {Eznaveh}}, \bibinfo {author} {\bibfnamefont {M.~A.}\ \bibnamefont {Eftekhar}}, \bibinfo {author} {\bibfnamefont {F.}~\bibnamefont {Wise}}, \bibinfo {author} {\bibfnamefont {R.~A.}\ \bibnamefont {Correa}},\ and\ \bibinfo {author} {\bibfnamefont {D.~N.}\ \bibnamefont {Christodoulides}},\ }\bibfield  {title} {\bibinfo {title} {A systematic analysis of parametric instabilities in nonlinear parabolic multimode fibers},\ }\bibfield  {journal} {\bibinfo  {journal} {APL Photonics}\ }\textbf {\bibinfo {volume} {4}},\ \href {https://doi.org/10.1063/1.5044659} {10.1063/1.5044659} (\bibinfo {year} {2019})\BibitemShut {NoStop}%
\bibitem [{\citenamefont {Bendahmane}\ \emph {et~al.}(2021)\citenamefont {Bendahmane}, \citenamefont {Conforti}, \citenamefont {Vanvincq}, \citenamefont {Arabí}, \citenamefont {Mussot},\ and\ \citenamefont {Kudlinski}}]{bendahmane2021}%
  \BibitemOpen
  \bibfield  {author} {\bibinfo {author} {\bibfnamefont {A.}~\bibnamefont {Bendahmane}}, \bibinfo {author} {\bibfnamefont {M.}~\bibnamefont {Conforti}}, \bibinfo {author} {\bibfnamefont {O.}~\bibnamefont {Vanvincq}}, \bibinfo {author} {\bibfnamefont {C.~M.}\ \bibnamefont {Arabí}}, \bibinfo {author} {\bibfnamefont {A.}~\bibnamefont {Mussot}},\ and\ \bibinfo {author} {\bibfnamefont {A.}~\bibnamefont {Kudlinski}},\ }\bibfield  {title} {\bibinfo {title} {Origin of spontaneous wave mixing processes in multimode grin fibers},\ }\href {https://doi.org/10.1364/oe.436229} {\bibfield  {journal} {\bibinfo  {journal} {Optics Express}\ }\textbf {\bibinfo {volume} {29}},\ \bibinfo {pages} {30822} (\bibinfo {year} {2021})}\BibitemShut {NoStop}%
\bibitem [{\citenamefont {Xia}\ \emph {et~al.}(2007)\citenamefont {Xia}, \citenamefont {Kumar}, \citenamefont {Cheng}, \citenamefont {Kulkarni}, \citenamefont {Islam}, \citenamefont {Galvanauskas}, \citenamefont {Terry}, \citenamefont {Freeman}, \citenamefont {Nolan},\ and\ \citenamefont {Wood}}]{xia2007}%
  \BibitemOpen
  \bibfield  {author} {\bibinfo {author} {\bibfnamefont {C.}~\bibnamefont {Xia}}, \bibinfo {author} {\bibfnamefont {M.}~\bibnamefont {Kumar}}, \bibinfo {author} {\bibfnamefont {M.-Y.}\ \bibnamefont {Cheng}}, \bibinfo {author} {\bibfnamefont {O.~P.}\ \bibnamefont {Kulkarni}}, \bibinfo {author} {\bibfnamefont {M.~N.}\ \bibnamefont {Islam}}, \bibinfo {author} {\bibfnamefont {A.}~\bibnamefont {Galvanauskas}}, \bibinfo {author} {\bibfnamefont {F.~L.}\ \bibnamefont {Terry}}, \bibinfo {author} {\bibfnamefont {M.~J.}\ \bibnamefont {Freeman}}, \bibinfo {author} {\bibfnamefont {D.~A.}\ \bibnamefont {Nolan}},\ and\ \bibinfo {author} {\bibfnamefont {W.~A.}\ \bibnamefont {Wood}},\ }\bibfield  {title} {\bibinfo {title} {Supercontinuum generation in silica fibers by amplified nanosecond laser diode pulses},\ }\href {https://doi.org/10.1109/JSTQE.2007.897414} {\bibfield  {journal} {\bibinfo  {journal} {IEEE Journal of Selected Topics in Quantum Electronics}\ }\textbf {\bibinfo {volume} {13}},\ \bibinfo {pages} {789}
  (\bibinfo {year} {2007})}\BibitemShut {NoStop}%
\bibitem [{\citenamefont {Yin}\ \emph {et~al.}(2016)\citenamefont {Yin}, \citenamefont {Zhang}, \citenamefont {Yao}, \citenamefont {Yang}, \citenamefont {Liu},\ and\ \citenamefont {Hou}}]{yin2016}%
  \BibitemOpen
  \bibfield  {author} {\bibinfo {author} {\bibfnamefont {K.}~\bibnamefont {Yin}}, \bibinfo {author} {\bibfnamefont {B.}~\bibnamefont {Zhang}}, \bibinfo {author} {\bibfnamefont {J.}~\bibnamefont {Yao}}, \bibinfo {author} {\bibfnamefont {L.}~\bibnamefont {Yang}}, \bibinfo {author} {\bibfnamefont {G.}~\bibnamefont {Liu}},\ and\ \bibinfo {author} {\bibfnamefont {J.}~\bibnamefont {Hou}},\ }\bibfield  {title} {\bibinfo {title} {1.9--3.6 $\mu$m supercontinuum generation in a very short highly nonlinear germania fiber with a high mid-infrared power ratio},\ }\href {https://doi.org/10.1364/OL.41.005067} {\bibfield  {journal} {\bibinfo  {journal} {Opt. Lett.}\ }\textbf {\bibinfo {volume} {41}},\ \bibinfo {pages} {5067} (\bibinfo {year} {2016})}\BibitemShut {NoStop}%
\bibitem [{\citenamefont {Jauregui}\ \emph {et~al.}(2012)\citenamefont {Jauregui}, \citenamefont {Steinmetz}, \citenamefont {Limpert},\ and\ \citenamefont {T\"{u}nnermann}}]{Jauregui2012}%
  \BibitemOpen
  \bibfield  {author} {\bibinfo {author} {\bibfnamefont {C.}~\bibnamefont {Jauregui}}, \bibinfo {author} {\bibfnamefont {A.}~\bibnamefont {Steinmetz}}, \bibinfo {author} {\bibfnamefont {J.}~\bibnamefont {Limpert}},\ and\ \bibinfo {author} {\bibfnamefont {A.}~\bibnamefont {T\"{u}nnermann}},\ }\bibfield  {title} {\bibinfo {title} {High-power efficient generation of visible and mid-infrared radiation exploiting four-wave-mixing in optical fibers},\ }\href {https://doi.org/10.1364/OE.20.024957} {\bibfield  {journal} {\bibinfo  {journal} {Opt. Express}\ }\textbf {\bibinfo {volume} {20}},\ \bibinfo {pages} {24957} (\bibinfo {year} {2012})}\BibitemShut {NoStop}%
\bibitem [{\citenamefont {Herzog}\ \emph {et~al.}(2012)\citenamefont {Herzog}, \citenamefont {Shamir},\ and\ \citenamefont {Ishaaya}}]{Herzog2012}%
  \BibitemOpen
  \bibfield  {author} {\bibinfo {author} {\bibfnamefont {A.}~\bibnamefont {Herzog}}, \bibinfo {author} {\bibfnamefont {A.}~\bibnamefont {Shamir}},\ and\ \bibinfo {author} {\bibfnamefont {A.~A.}\ \bibnamefont {Ishaaya}},\ }\bibfield  {title} {\bibinfo {title} {Wavelength conversion of nanosecond pulses to the mid-ir in photonic crystal fibers},\ }\href {https://doi.org/10.1364/OL.37.000082} {\bibfield  {journal} {\bibinfo  {journal} {Opt. Lett.}\ }\textbf {\bibinfo {volume} {37}},\ \bibinfo {pages} {82} (\bibinfo {year} {2012})}\BibitemShut {NoStop}%
\bibitem [{\citenamefont {Dupiol}\ \emph {et~al.}(2017{\natexlab{a}})\citenamefont {Dupiol}, \citenamefont {Bendahmane}, \citenamefont {Krupa}, \citenamefont {Tonello}, \citenamefont {Fabert}, \citenamefont {Kibler}, \citenamefont {Sylvestre}, \citenamefont {Barthelemy}, \citenamefont {Couderc}, \citenamefont {Wabnitz},\ and\ \citenamefont {Millot}}]{dupiol2017}%
  \BibitemOpen
  \bibfield  {author} {\bibinfo {author} {\bibfnamefont {R.}~\bibnamefont {Dupiol}}, \bibinfo {author} {\bibfnamefont {A.}~\bibnamefont {Bendahmane}}, \bibinfo {author} {\bibfnamefont {K.}~\bibnamefont {Krupa}}, \bibinfo {author} {\bibfnamefont {A.}~\bibnamefont {Tonello}}, \bibinfo {author} {\bibfnamefont {M.}~\bibnamefont {Fabert}}, \bibinfo {author} {\bibfnamefont {B.}~\bibnamefont {Kibler}}, \bibinfo {author} {\bibfnamefont {T.}~\bibnamefont {Sylvestre}}, \bibinfo {author} {\bibfnamefont {A.}~\bibnamefont {Barthelemy}}, \bibinfo {author} {\bibfnamefont {V.}~\bibnamefont {Couderc}}, \bibinfo {author} {\bibfnamefont {S.}~\bibnamefont {Wabnitz}},\ and\ \bibinfo {author} {\bibfnamefont {G.}~\bibnamefont {Millot}},\ }\bibfield  {title} {\bibinfo {title} {Far-detuned cascaded intermodal four-wave mixing in a multimode fiber},\ }\href {https://doi.org/10.1364/ol.42.001293} {\bibfield  {journal} {\bibinfo  {journal} {Optics Letters}\ }\textbf {\bibinfo {volume} {42}},\ \bibinfo {pages} {1293} (\bibinfo {year}
  {2017}{\natexlab{a}})}\BibitemShut {NoStop}%
\bibitem [{\citenamefont {Sunak}\ and\ \citenamefont {Bastien}(1989)}]{sunak1989}%
  \BibitemOpen
  \bibfield  {author} {\bibinfo {author} {\bibfnamefont {H.}~\bibnamefont {Sunak}}\ and\ \bibinfo {author} {\bibfnamefont {S.}~\bibnamefont {Bastien}},\ }\bibfield  {title} {\bibinfo {title} {Refractive index and material dispersion interpolation of doped silica in the 0.6-1.8 micrometer wavelength region},\ }\href {https://doi.org/10.1109/68.36016} {\bibfield  {journal} {\bibinfo  {journal} {IEEE Photonics Technology Letters}\ }\textbf {\bibinfo {volume} {1}},\ \bibinfo {pages} {142} (\bibinfo {year} {1989})}\BibitemShut {NoStop}%
\bibitem [{\citenamefont {Dupiol}\ \emph {et~al.}(2017{\natexlab{b}})\citenamefont {Dupiol}, \citenamefont {Bendahmane}, \citenamefont {Krupa}, \citenamefont {Fatome}, \citenamefont {Tonello}, \citenamefont {Fabert}, \citenamefont {Couderc}, \citenamefont {Wabnitz},\ and\ \citenamefont {Millot}}]{dupiol2017immi}%
  \BibitemOpen
  \bibfield  {author} {\bibinfo {author} {\bibfnamefont {R.}~\bibnamefont {Dupiol}}, \bibinfo {author} {\bibfnamefont {A.}~\bibnamefont {Bendahmane}}, \bibinfo {author} {\bibfnamefont {K.}~\bibnamefont {Krupa}}, \bibinfo {author} {\bibfnamefont {J.}~\bibnamefont {Fatome}}, \bibinfo {author} {\bibfnamefont {A.}~\bibnamefont {Tonello}}, \bibinfo {author} {\bibfnamefont {M.}~\bibnamefont {Fabert}}, \bibinfo {author} {\bibfnamefont {V.}~\bibnamefont {Couderc}}, \bibinfo {author} {\bibfnamefont {S.}~\bibnamefont {Wabnitz}},\ and\ \bibinfo {author} {\bibfnamefont {G.}~\bibnamefont {Millot}},\ }\bibfield  {title} {\bibinfo {title} {Intermodal modulational instability in graded-index multimode optical fibers},\ }\href {https://doi.org/10.1364/OL.42.003419} {\bibfield  {journal} {\bibinfo  {journal} {Opt. Lett.}\ }\textbf {\bibinfo {volume} {42}},\ \bibinfo {pages} {3419} (\bibinfo {year} {2017}{\natexlab{b}})}\BibitemShut {NoStop}%
\bibitem [{\citenamefont {Stolen}(1975)}]{stolen1975}%
  \BibitemOpen
  \bibfield  {author} {\bibinfo {author} {\bibfnamefont {R.}~\bibnamefont {Stolen}},\ }\bibfield  {title} {\bibinfo {title} {Phase-matched-stimulated four-photon mixing in silica-fiber waveguides},\ }\href {https://doi.org/10.1109/JQE.1975.1068571} {\bibfield  {journal} {\bibinfo  {journal} {IEEE Journal of Quantum Electronics}\ }\textbf {\bibinfo {volume} {11}},\ \bibinfo {pages} {100} (\bibinfo {year} {1975})}\BibitemShut {NoStop}%
\bibitem [{\citenamefont {Agrawal}(2019)}]{agrawalbook}%
  \BibitemOpen
  \bibfield  {author} {\bibinfo {author} {\bibfnamefont {G.~P.}\ \bibnamefont {Agrawal}},\ }\href@noop {} {\emph {\bibinfo {title} {Nonlinear {Fiber} {Optics} {6th ed.}}}}\ (\bibinfo  {publisher} {Academic Press},\ \bibinfo {year} {2019})\BibitemShut {NoStop}%
\bibitem [{\citenamefont {B\'ejot}(2019)}]{bejot2019multimodal}%
  \BibitemOpen
  \bibfield  {author} {\bibinfo {author} {\bibfnamefont {P.}~\bibnamefont {B\'ejot}},\ }\bibfield  {title} {\bibinfo {title} {Multimodal unidirectional pulse propagation equation},\ }\href {https://doi.org/10.1103/PhysRevE.99.032217} {\bibfield  {journal} {\bibinfo  {journal} {Phys. Rev. E}\ }\textbf {\bibinfo {volume} {99}},\ \bibinfo {pages} {032217} (\bibinfo {year} {2019})}\BibitemShut {NoStop}%
\bibitem [{\citenamefont {Harvey}\ \emph {et~al.}(2003)\citenamefont {Harvey}, \citenamefont {Leonhardt}, \citenamefont {Coen}, \citenamefont {Wong}, \citenamefont {Knight}, \citenamefont {Wadsworth},\ and\ \citenamefont {Russell}}]{harvey2003}%
  \BibitemOpen
  \bibfield  {author} {\bibinfo {author} {\bibfnamefont {J.~D.}\ \bibnamefont {Harvey}}, \bibinfo {author} {\bibfnamefont {R.}~\bibnamefont {Leonhardt}}, \bibinfo {author} {\bibfnamefont {S.}~\bibnamefont {Coen}}, \bibinfo {author} {\bibfnamefont {G.~K.~L.}\ \bibnamefont {Wong}}, \bibinfo {author} {\bibfnamefont {J.~C.}\ \bibnamefont {Knight}}, \bibinfo {author} {\bibfnamefont {W.~J.}\ \bibnamefont {Wadsworth}},\ and\ \bibinfo {author} {\bibfnamefont {P.~S.~J.}\ \bibnamefont {Russell}},\ }\bibfield  {title} {\bibinfo {title} {Scalar modulation instability in the normal dispersion regime by use of a photonic crystal fiber},\ }\href {https://doi.org/10.1364/OL.28.002225} {\bibfield  {journal} {\bibinfo  {journal} {Optics Letters}\ }\textbf {\bibinfo {volume} {28}},\ \bibinfo {pages} {2225} (\bibinfo {year} {2003})}\BibitemShut {NoStop}%
\bibitem [{\citenamefont {Godin}\ \emph {et~al.}(2014)\citenamefont {Godin}, \citenamefont {Combes}, \citenamefont {Ahmad}, \citenamefont {Rochette}, \citenamefont {Sylvestre},\ and\ \citenamefont {Dudley}}]{godin2014}%
  \BibitemOpen
  \bibfield  {author} {\bibinfo {author} {\bibfnamefont {T.}~\bibnamefont {Godin}}, \bibinfo {author} {\bibfnamefont {Y.}~\bibnamefont {Combes}}, \bibinfo {author} {\bibfnamefont {R.}~\bibnamefont {Ahmad}}, \bibinfo {author} {\bibfnamefont {M.}~\bibnamefont {Rochette}}, \bibinfo {author} {\bibfnamefont {T.}~\bibnamefont {Sylvestre}},\ and\ \bibinfo {author} {\bibfnamefont {J.~M.}\ \bibnamefont {Dudley}},\ }\bibfield  {title} {\bibinfo {title} {Far-detuned mid-infrared frequency conversion via normal dispersion modulation instability in chalcogenide microwires},\ }\href {https://doi.org/10.1364/OL.39.001885} {\bibfield  {journal} {\bibinfo  {journal} {Optics Letters}\ }\textbf {\bibinfo {volume} {39}},\ \bibinfo {pages} {1885} (\bibinfo {year} {2014})}\BibitemShut {NoStop}%
\bibitem [{\citenamefont {Leventoux}\ \emph {et~al.}(2021)\citenamefont {Leventoux}, \citenamefont {Granger}, \citenamefont {Krupa}, \citenamefont {Mansuryan}, \citenamefont {Fabert}, \citenamefont {Tonello}, \citenamefont {Wabnitz}, \citenamefont {Couderc},\ and\ \citenamefont {Février}}]{leventoux2021}%
  \BibitemOpen
  \bibfield  {author} {\bibinfo {author} {\bibfnamefont {Y.}~\bibnamefont {Leventoux}}, \bibinfo {author} {\bibfnamefont {G.}~\bibnamefont {Granger}}, \bibinfo {author} {\bibfnamefont {K.}~\bibnamefont {Krupa}}, \bibinfo {author} {\bibfnamefont {T.}~\bibnamefont {Mansuryan}}, \bibinfo {author} {\bibfnamefont {M.}~\bibnamefont {Fabert}}, \bibinfo {author} {\bibfnamefont {A.}~\bibnamefont {Tonello}}, \bibinfo {author} {\bibfnamefont {S.}~\bibnamefont {Wabnitz}}, \bibinfo {author} {\bibfnamefont {V.}~\bibnamefont {Couderc}},\ and\ \bibinfo {author} {\bibfnamefont {S.}~\bibnamefont {Février}},\ }\bibfield  {title} {\bibinfo {title} {Frequency-resolved spatial beam mapping in multimode fibers: application to mid-infrared supercontinuum generation},\ }\href {https://doi.org/10.1364/ol.428623} {\bibfield  {journal} {\bibinfo  {journal} {Optics Letters}\ }\textbf {\bibinfo {volume} {46}},\ \bibinfo {pages} {3717} (\bibinfo {year} {2021})}\BibitemShut {NoStop}%
\end{thebibliography}%

\end{document}